\documentclass[twocolumn,twoside,slac_two]{revtex4}
\usepackage{graphicx}
\usepackage{fancyhdr}
\def\fermi{\textit{Fermi}}

\def\lesssim{\lower4pt\hbox{${\buildrel < \over \sim}$}}
\def\gtrsim{\lower4pt\hbox{${\buildrel > \over \sim}$}}

\begin{document}

\title{Modeling the Spectral Energy Distributions
and Variability of Blazars}


\author{M. Boettcher}
\affiliation{Astrophysical Institute, Department of Physics and Astronomy,
Ohio University, Athens, OH 45701, USA}

\begin{abstract}
In this review, recent progress in theoretical models for 
the broadband (radio through $\gamma$-ray) emission from blazars are summarized. 
The salient features of both leptonic and hadronic models are reviewed.
I present sample modeling results of spectral energy distributions (SEDs)
of different types of {\it Fermi}-detected blazars along the traditional
blazar sequence, using both types of models. In many cases,
the SEDs of high-frequency peaked blazars (HBLs) have been found to be well 
represented by simple synchrotron + synchrotron self-Compton (SSC) models. 
However, a few HBLs recently discovered as very-high-energy (VHE) gamma-ray
emitters by VERITAS are actually better represented by either external-Compton
or hadronic models. Often, spectral modeling with time-independent single-zone 
models alone is not sufficient to constrain models, as both leptonic and 
lepto-hadronic models are able to provide acceptable fits to the overall 
SED. This degeneracy can be lifted by considering further constraints from 
spectral variability. Recent developments of time-dependent and inhomogeneous 
blazar models will be discussed, including detailed numerical simulations 
as well as a semi-analytical approach to the time-dependent radiation 
signatures of shock-in-jet models.
\end{abstract}

\maketitle

\thispagestyle{fancy}


\section{\label{intro}INTRODUCTION}

\begin{figure*}
\centering
\includegraphics[width=0.95\textwidth]{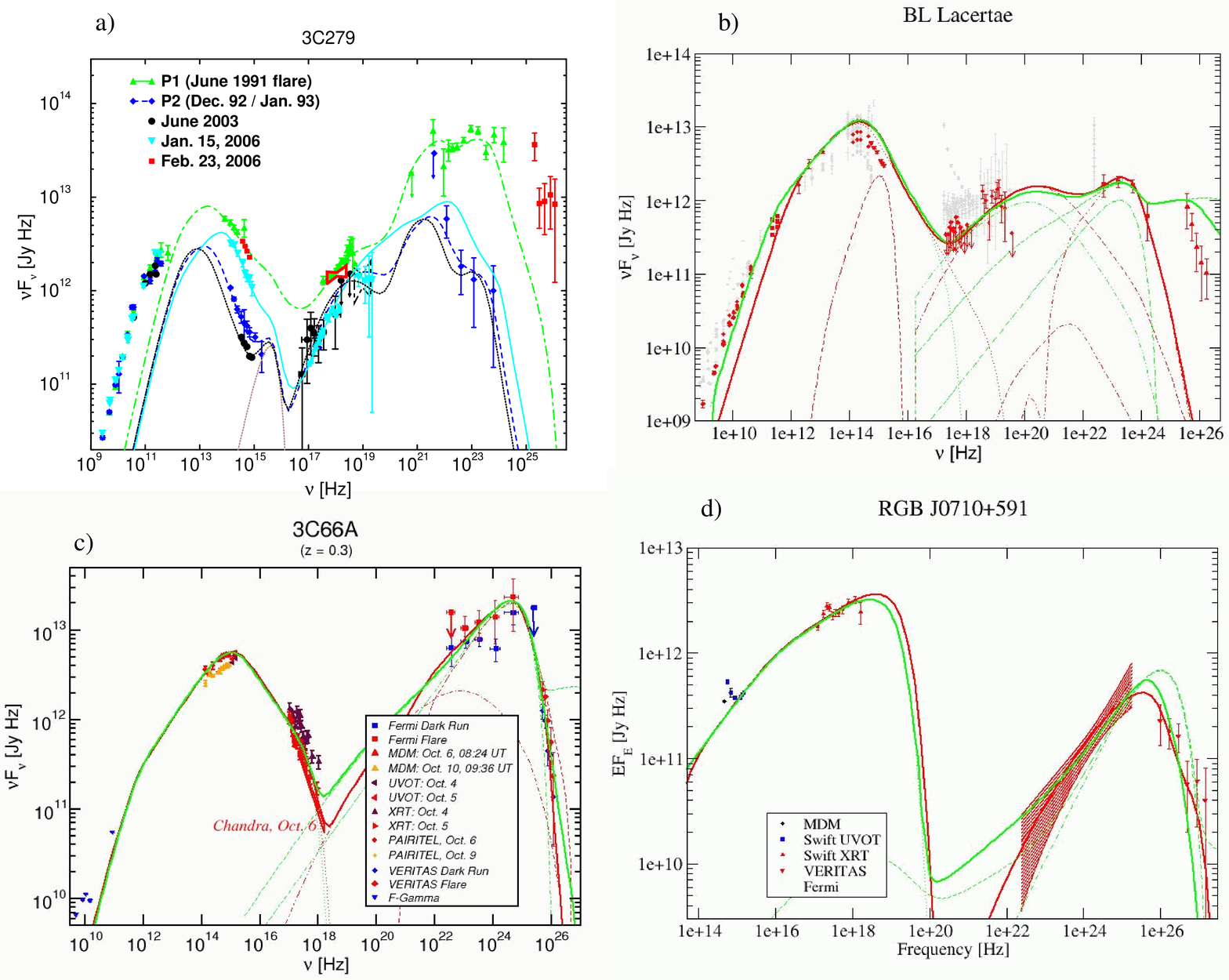}
\caption{\label{SEDs}Spectral energy distributions of four sub-classes
of blazars: a) the FSRQ 3C279 (from \cite{collmar10}), b) the LBL BL Lacertae,
(data from \cite{abdo10b}), c) the intermediate BL Lac 3C66A (data from
\cite{abdo11a}), and d) the HBL RGB J0710+591 (data from \cite{acciari10b}).
In Panel a) (3C279), lines are one-zone leptonic model fits to SEDs at
various epochs shown in the figure. In all other panels, red lines
are fits with a leptonic one-zone model; green lines are fits
with a one-zone lepto-hadronic model. }
\end{figure*}

Blazars (BL~Lac objects and $\gamma$-ray loud flat spectrum
radio quasars [FSRQs]) are the most extreme class of active
galaxies known. Their broadband (from radio to $\gamma$-rays) 
continuum SEDs are dominated by non-thermal emission and consist
of two distinct, broad components: a low-energy component
from radio through UV or X-rays, and a high-energy component
from X-rays to $\gamma$-rays (see, e.g., Figure \ref{SEDs}).

Blazars are sub-divided into several types, defined by the location
of the peak of the low-energy (synchrotron) SED component. Low-synchrotron-peaked
(LSP) blazars, consisting of flat-spectrum radio quasars (FSRQs) and low-frequency
peaked BL Lac objects (LBLs), have their synchrotron peak in the infrared
regime, at $\nu_s \le 10^{14}$~Hz. Intermediate-Synchrotron-Peaked (ISP)
blazars, including LBLs and intermediate BL Lac objects (IBLs), have their 
synchrotron peak at optical -- near-UV frequencies at $10^{14} \, {\rm Hz}
< \nu_s \le 10^{15}$~Hz, while High-Synchrotron-Peaked (HSP) blazars,
which are almost all high-frequency-peaked BL Lac objects (HBL), have
their synchrotron peak at X-ray energies with $\nu_s > 10^{15}$~Hz
\citep{abdo10b}. Such a sequence appears to be more physical than
the historical distinction between quasars and BL~Lac objects based
on the equivalent width of optical emission lines \citep{landt04}, 
as the equivalent width is dependent on the strongly time-variable
flux of the non-thermal continuum \citep{giommi12}. 

The blazar sequence had first been identified by \cite{fossati98}, and 
associated also with a trend of overall decreasing bolometric luminosity
as well as decreasing $\gamma$-ray dominance along the sequence FSRQ $\to$
LBL $\to$ HBL. According to this classification, the bolometric power output
of FSRQs is strongly $\gamma$-ray dominated, especially during flaring 
states, while HBLs are expected to be always synchrotron dominated.
However, while the overall bolometric-luminosity trend still seems to hold,
recently, even HBLs seem to undergo episodes of strong $\gamma$-ray dominance.
A prominent example was the giant VHE $\gamma$-ray flare of the HBL PKS~2155-304 
in 2006, during which the SED was strongly $\gamma$-ray dominated 
\citep{aharonian09}.

Figure \ref{SEDs} shows examples of blazar SEDs along the blazar
sequence, from the FSRQ 3C279 (a), via the LBL BL~Lacertae (b) and
the IBL 3C~66A (c), to the HBL RGB J0710+591 (d). The sequence of increasing
synchrotron peak frequency is clearly visible. However, the \fermi\
spectrum of the LBL BL~Lacertae indicates a $\gamma$-ray flux clearly
below the synchrotron level, while the SED of the IBL 3C~66A is clearly
dominated by the \fermi\ $\gamma$-ray flux, in contradiction with the
traditional blazar sequence.

Blazars are known to be variable at all wavelengths. Typically, the 
variability amplitudes are the largest and variability time scales 
are the shortest at the high-frequency ends of the two SED components. 
In HBLs, this refers to the X-ray and VHE $\gamma$-ray regimes. 
High-energy variability in some HBLs (PKS~2155-305, Mrk~501, Mrk~421) 
has been observed on time scales down to just a few minutes 
\citep{aharonian07,albert07a,fortson11}. The flux variability of 
blazars is often accompanied by spectral changes which are 
sometimes associated with inter-band or intra-band time lags.
In a few HBLs (especially, Mrk~421), variability patterns have 
been found which can be characterized as spectral hysteresis in 
hardness-intensity diagrams
\citep[e.g.,][]{takahashi96,kataoka00,fossati00,zhang02}. However,
even within the same object this feature tends not to be persistent
over multiple observations. Also in other types of blazars, hints
of time lags between different observing bands are occasionally
found in individual observing campaigns (e.g., \cite{boettcheral07,horan09}),
but the search for time-lag patterns persisting throughout multiple
years has so far remained unsuccessful \citep[see, e.g.,][for a
systematic search for time lags between optical, X-ray and $\gamma$-ray
emission in the quasar 3C279]{hartman01}.

\section{\label{models}LEPTONIC AND HADRONIC BLAZAR MODELS}

The high apparent bolometric luminosities (assuming isotropic
emission), rapid variability, and apparent superluminal motions 
of radio components in blazar jets provide compelling evidence
that the nonthermal continuum emission of blazars is produced
in $\lesssim$~1 light day sized emission regions, propagating
relativistically with velocity $\beta_{\Gamma}$c along a jet
directed at a small angle $\theta_{\rm obs}$ with respect to
our line of sight \citep[e.g.,][]{savolainen10}. Let $\Gamma =
(1 - \beta_{\Gamma}^2)^{-1/2}$ be the bulk Lorentz factor of
the  emission region, then Doppler boosting is determined by the Doppler 
factor $D = (\Gamma [1 - \beta_{\Gamma} \cos\theta_{\rm obs}])^{-1}$.
In the following, primes denote quantities in the co-moving frame of 
the emission region. The observed frequency $\nu_{\rm obs}$ is 
related to the emitted frequency through $\nu_{\rm obs} = D \, \nu' / 
(1 + z)$, where $z$ is the redshift of the source. The energy fluxes 
are connected through $F_{\nu_{\rm obs}}^{\rm obs} = D^3 \, F'_{\nu'}$.
Intrinsic variability on a co-moving time scale $t'_{\rm var}$ will
be observed on a time scale $t_{\rm var}^{\rm obs} = t'_{\rm var} \,
(1 + z) / D$. Using the latter transformation along with causality
arguments, any observed variability leads to an upper limit on the
size scale of the emission region through $R \lesssim c \, 
t_{\rm var}^{\rm obs} \, D / (1 + z)$.

It is generally agreed that the low-frequency component in the blazar 
SED arises from synchrotron emission of relativistic electrons. 
However, there are two fundamentally different approaches concerning 
the high-energy emission. If protons are not accelerated to ultrarelativistic
energies (exceeding the threshold for $p\gamma$ pion production on 
synchrotron and/or external photons), the high-energy radiation will 
be dominated by Compton emission from ultrarelativistic electrons and/or 
pairs (leptonic models). In the opposite case, the high-energy emission 
will be dominated by cascades initiated by $p\gamma$ pair and pion production 
as well as proton, $\pi^{\pm}$, and $\mu^{\pm}$ synchrotron radiation, while 
primary leptons are still responsible for the low-frequency synchrotron
emission (hadronic models). The following sub-sections provide a brief 
overview of the main radiation physics aspects of both leptonic and 
hadronic models. A more in-depth review of the radiation physics
in relativistic jets can be found in \cite{br12}.

When fitting SEDs of blazars, consideration should be given to the
energy balance between the magnetic field and the particle content
in the jet, as this contains information on the jet launching and
acceleration mechanisms. If the relativistic jets of AGN are powered
by the rotational energy of the central black hole \citep{bz77},
the jets are expected to be initially Poynting-flux dominated.
The energy thus carried primarily in the form of magnetic fields
then needs to transferred to relativistic particles which can then
produce the observed high-energy emission. This energy conversion
is expected to cease when the energy densities in the magnetic field
and in relativistic particles approach equipartition. In this scenario,
the jets are therefore not expected to become matter dominated within
the central few parsecs of AGN, where the high-energy emission in
blazars is believed to be produced \citep{lyubarsky11}. Also, if
magnetic pressure plays an essential role in collimating and confining
the jets out to pc or kpc scales, the energy density (and hence pressure)
of particles in the jet can not dominate over the magnetic-field
pressure since otherwise the jets would simply expand conically. 

If SED fits require far sub-equipartition magnetic fields, this may
indicate that the jets are magneto-hydrodynamically powered by the 
accretion flow. In this case, magnetic fields may be self-generated 
in shear flows in the case of a fast inner spine surrounded by a slower, 
mildly relativistic outer flow (sheath). Kelvin-Helmholtz type instabilities 
can then lead to anisotropic particle acceleration, and the self-generated 
magnetic fields are expected to remain below equipartition with the
thermal energy carried in the flow \citep{liang11,alves11}.

\subsection{\label{leptonic}Leptonic models}

In leptonic models, the high-energy emission is produced via Compton
upscattering of soft photons off the same ultrarelativistic electrons
which are producing the synchrotron emission. Both the synchrotron 
photons produced within the jet \citep[SSC = Synchrotron 
Self-Compton:][]{mg85,maraschi92,bm96}, and external 
photons (EC = External Compton) can serve as target photons 
for Compton scattering. Possible sources of external seed photons 
include the accretion disk radiation \citep[e.g.,][]{dermer92,ds93}, 
reprocessed optical -- UV emission from circumnuclear material 
\citep[e.g., the BLR;][]{sikora94,bl95,gm96,dermer97},
infrared emission from warm dust \citep{blazejowski00}, or
synchrotron emission from other (faster/slower) regions of the
jet itself \citep{gk03,gt08}.

If the observed blazar emission were arising in a stationary
emission region (whose size is constrained by causality arguments),
its compactness would, in many cases, be too large for high-energy 
$\gamma$-rays to escape from it. This problem is overcome by 
relativistic Doppler boosting so that model fits can be achieved
with parameters such that the $\gamma\gamma$ opacity of the emission 
region is low throughout most of the high-energy spectrum. However, 
at the highest photon energies, this effect may make a non-negligible 
contribution to the formation of the emerging spectrum \citep{aharonian08} 
and lead to the re-processing of some of the radiated power towards 
lower frequencies. $\gamma\gamma$ absorption within the AGN may also
occur on photons external to the emission region, as required to
be present if $\gamma$-rays are produced through the EC mechanism
\citep[e.g.][]{reimer07,liu08,ps10}. This absorption process might
lead to Compton supported pair cascades which are expected to release
their peak power at MeV -- GeV energies \citep[e.g.,][]{rb10,rb11}.

Also the deceleration of the jets may have a significant impact on 
the observable properties of blazar emission through the radiative 
interaction of emission regions with different speed \citep{gk03,ghisellini05}.
The deceleration of the high-energy emission region is expected to
cause characteristic variability features which may be reminiscent
of those seen in gamma-ray burst light curves \citep{bp09}. Varying 
Doppler factors may also be a result of a slight change in the jet 
orientation without a substantial change in speed, e.g., in a helical-jet 
configuration \cite[e.g.,][]{vr99}. In the case of ordered magnetic-field 
structures in the emission region, such a helical configuration should 
have observable synchrotron polarization signatures, such as the
prominent polarization-angle swing recently observed in conjunction
with an optical + \fermi\ $\gamma$-ray flare of 3C~279 \citep{abdo10a}.

In the simplest models, the underlying lepton (electrons
and/or positrons) distribution is ad-hoc pre-specified, either as
a single or broken power-law with a low- and high-energy cut-off.
While leptonic models under this assumption have been successful
in modeling blazar SEDs \citep[e.g.,][]{ghisellini98}, they lack 
a self-consistent basis for the shape of the electron distribution.
A more realistic approach consists of the self-consistent steady-state
solution of the Fokker-Planck equation governing the evolution of
the electron distribution function, including a physically motivated
(e.g., from shock-acceleration theory) acceleration of particles and
all relevant radiative and adiabatic cooling mechanisms 
\citep[e.g.,][]{gt09,acciari09b,weidinger10}. The model described 
in \cite{acciari09b}, based on the time-dependent model of \cite{bc02}
\citep[see also][]{boettcher12}, has been used to produce the leptonic 
model fits shown in Fig. \ref{SEDs}.

In order to reproduce not only broadband SEDs, but also variability
patterns, the time-dependent electron dynamics and radiation transfer
problem has to be solved self-consistently. Such time-dependent SSC
models have been developed by, e.g., \cite{mk97,kataoka00,lk00,sokolov04}.
External radiation fields have been included in such treatments in, e.g.,
\cite{sikora01,bc02,sm05}.

Leptonic models have generally been very successfully applied to model the 
SEDs and (in few cases) spectral variability of blazars. The radiative cooling
time scales (in the observers's frame) of synchrotron-emitting electrons
in a typical $B \sim 1$~G magnetic field are of the order of several hours
-- $\sim 1$~d at optical frequencies and $\lesssim 1$~hr in X-rays
and hence compatible with the observed intra-day variability. However,
the recent observation of extremely rapid VHE $\gamma$-ray variability
on time scales of a few minutes poses severe problems to simple one-zone
leptonic emission models. Even with large bulk Lorentz factors of
$\sim 50$, causality requires a size of the emitting region that might
be smaller than the Schwarzschild radius of the central black hole of
the AGN \citep{begelman08}. As a possible solution, it has been
suggested \citep{tg08} that the $\gamma$-ray emission region may, in
fact, be only a small spine of ultrarelativistic plasma within a larger,
slower-moving jet. Such fast-moving small-scale jets could plausibly
be powered by magnetic reconnection in a Poynting-flux dominated jet,
as proposed by \cite{giannios09}.

\subsection{\label{hadronic}Hadronic models}

If a significant fraction of the jet power is converted into the
acceleration of relativistic protons in a strongly magnetized
environment, reaching the threshold for $p\gamma$ pion production,
synchrotron-supported pair cascades will develop, initiated by
primary $\pi^0$ decay photons and synchrotron emission from
secondary pions, muons and electrons/positrons at 
ultra-high $\gamma$-ray energies, where the emission region is
highly opaque to $\gamma\gamma$ absorption \citep{mb92,mannheim93}.
The acceleration of protons to the necessary ultrarelativistic energies
($E_p^{\rm max} \gtrsim 10^{19}$~eV) requires high magnetic fields of
several tens of Gauss to constrain the Larmor radius $R_L = 3.3 \times
10^{15} \, B_1^{-1} \, E_{19}$~cm, where $B = 10 \, B_1$~G, and $E_p
= 10^{19} \, E_{19}$~eV, to be smaller than the size of the
emission region (typically inferred to be $R \lesssim 10^{16}$~cm from
the observed variability time scale). In the presence of such high magnetic
fields, the synchrotron radiation of the primary protons \citep{aharonian00,mp00}
and of secondary muons and mesons \citep{rm98,mp00,mp01,muecke03}
must be taken into account in order to construct a self-consistent
synchrotron-proton blazar (SPB) model. Electromagnetic cascades can be
initiated by photons from $\pi^0$-decay (``$\pi^0$ cascade''), electrons
from the $\pi^\pm\to \mu^\pm\to e^\pm$ decay (``$\pi^\pm$ cascade''),
$p$-synchrotron photons (``$p$-synchrotron cascade''), and $\mu$-, $\pi$-
and $K$-synchrotron photons (``$\mu^\pm$-synchrotron cascade'').

\cite{mp01} and \cite{muecke03} have shown that the ``$\pi^0$ cascades'' 
and ``$\pi^\pm$ cascades'' from ultra-high energy protons generate 
featureless $\gamma$-ray spectra, in contrast to ``$p$-synchrotron 
cascades'' and``$\mu^\pm$-synchrotron cascades'' that produce a 
two-component $\gamma$-ray spectrum. In general, direct proton 
and $\mu^{\pm}$ synchrotron radiation is mainly responsible for the 
high energy bump in blazars, whereas the low energy bump is dominanted
by synchrotron radiation from the primary $e^-$, with a contribution from
secondary electrons.

Hadronic blazar models have so far been very difficult to investigate in a
time-dependent way because of the very time-consuming nature of the required
Monte-Carlo cascade simulations. In general, it appears that it is difficult
to reconcile very rapid high-energy variability with the radiative cooling
time scales of protons, e.g., due to synchrotron emission, which is
$t_{\rm sy}^{\rm obs} = 4.5 \times 10^5 \, (1 + z) \, D_1^{-1} B_1^{-2}
\, E_{19}^{-1}$~s \citep{aharonian00}, i.e., of the order of several
days for $\sim 10$~G magnetic fields and typical Doppler factors
$D = 10 \, D_1$. However, rapid variability on time scales shorter
than the proton cooling time scale may be caused by geometrical effects,
for example in a helical jet.

In order to avoid time-consuming Monte-Carlo simulations of the hadronic
processes and cascades involved in lepto-hadronic models, one may
consider a simplified prescription of the hadronic processes.
\cite{ka08} have produced analytic fit functions to Monte-Carlo
generated results of hadronic interactions using the SOFIA code
\citep{muecke00}. Those fits describe the spectra of the final
decay products, such as electrons, positrons, neutrinos, and
photons from $\pi^0$ decay. This neglects the possibility of
substantial synchrotron losses to muons and pions before they
decay. Since the muon life time is longer than the pion life
time, the more stringent condition is given by the requirement
that $t_{{\rm sy},\mu} \gg t_{{\rm d},\mu}$, where $t_{\rm sy}$
and $t_{\rm d}$ denote the synchrotron and decay time scales,
respectively. This can be translated into the requirement that
$B \ll 5.6 \gamma_{\rm p, max, 10}^{-1}$~G, where $\gamma_{\rm p,
max, 10}$ is the Lorentz factor of the highest-energy protons
in units of $10^{10}$. \cite{huemmer10} have developed analytical
fits to the secondary-production spectra in hadronic processes
for all individual steps, thus allowing one to take muon and
pion synchrotron (and Compton) emission into account properly.

A hadronic blazar model based on the templates of \cite{ka08} 
has been developed by \cite{boettcher12} \citep[see also][]{boettcher10}. 
This model employs a semi-analytical solution for UHE $\gamma$-ray 
induced, synchrotron-supported pair cascades. The hadronic
fits shown in Fig \ref{SEDs}, b) -- d) have been performed
with this model. As the low-frequency component is electron-synchrotron 
emission from primary electrons, it is not surprising that virtually 
identical fits to the synchrotron component can be provided in both 
types of models. In the high-frequency component, strongly peaked
spectral shapes, as, e.g., in 3C~66A and RGB J0710+591 require a strong
proton-synchrotron dominance with the cascading of higher-energy ($\gg$~TeV)
emission only making a minor contribution to the high-energy emission.
This, in fact, makes it difficult to achieve a substantial extension of
the escaping high-energy emission into the $> 100$~GeV VHE $\gamma$-ray
regime. In objects with a smoother high-energy SED, e.g., BL Lacertae
in Fig. \ref{SEDs}b, a substantially larger contribution from cascade
emission (and leptonic SSC emission) is allowed to account for a
relatively high level of hard X-ray / soft $\gamma$-ray emission. This
also allows for a substantial extension of the $\gamma$-ray spectrum
into the VHE regime.

\section{\label{fits}SED FITS ALONG THE BLAZAR SEQUENCE}

In the framework of leptonic models, the blazar sequence LSP $\to$
ISP $\to$ HSP is often modeled through a decreasing contribution
of external radiation fields to radiative cooling of electrons and
associated EC $\gamma$-ray emission \citep{ghisellini98}. In this sense,
HBLs have traditionally been well represented by pure synchrotron-self-Compton
models, while FSRQs usually require a substantial EC component. This
interpretation is consistent with the observed strong emission lines
in FSRQs, which are absent in BL~Lac objects. At the same time, the
denser circumnuclear environment in quasars might also lead to a
higher accretion rate and hence a more powerful jet, consistent with
the overall trend of bolometric luminosities along the blazar sequence.
This may even be related to an evolutionary sequence from FSRQs to HBLs
governed by the gradual depletion of the circumnuclear environment
\citep{bd02}.

However, in this interpretation, it would be expected that mostly HBLs 
(and maybe IBLs) should be detectable as emitters of VHE $\gamma$-rays
since in LBLs and FSRQs, electrons are not expected to reach $\sim$~TeV
energies. This paradigm has been undermined by the recent VHE $\gamma$-ray 
detections of lower-frequency peaked objects such as W~Comae \citep{acciari08}, 
3C66A \citep{acciari09a}, PKS~1424+240 \citep{acciari10a}, BL~Lacertae
\citep{albert07b}, S5~0716+714 \citep{anderhub09}, and even the FSRQs
3C~279 \citep{albert08}, PKS~1222+21 \citep{aleksic11} and PKS~1510-089 
\citep{wb10}.

The overall SEDs of IBLs detected by VERITAS could still be fit
satisfactorily with a purely leptonic model.  Fitting the SEDs of
the IBLs 3C66A and W~Comae with a pure SSC model, while formally
possible, would require rather extreme parameters. In particular,
magnetic fields several orders of magnitude below equipartition
would be needed, which might pose a severe problem for jet collimation
(see the discussion in Section \ref{models}).
Much more natural fit parameters can be adopted when including an EC
component with an infrared radiation field as target photons
\citep{acciari09b,abdo11a}.

In \cite{boettcher09} it has been demonstrated that the VHE $\gamma$-ray 
detection of the FSRQ 3C~279 poses severe problems for any variation of 
a single-zone leptonic jet model. The fundamental
problem for the leptonic model interpretation is the extremely wide
separation between the synchrotron and $\gamma$-ray peak frequencies.
In a single-zone SSC interpretation this would require a very high
Lorentz factor of electrons at the peak of the electron distribution
and, in turn, an extremely low magnetic field. In an EC interpretation
the SSC component, usually dominating the X-ray emission in leptonic 
fits to FSRQs like 3C~279, is too far suppressed to model the 
simultaneous RXTE spectrum. As an alternative, the optical emission 
could be produced in a separate emission component. A pure SSC fit 
to the X-ray and $\gamma$-ray component is technically possible, 
but also requires a far sub-equipartition magnetic field. Much 
more natural parameters could be achieved in a fit with the 
synchrotron-proton-blazar model of \cite{muecke03}.

\subsection{\label{noSSC}HBLS disfavouring SSC}

Until very recently, the SEDs of essentially all HBLs detected in
VHE $\gamma$-rays could be well modelled using simple leptonic SSC 
models. The most attractive feature of this model is its relative 
simplicity, requiring only a small number of fit parameters. If
the peak frequencies and corresponding fluxes in a blazar SED are
well measured and an additional constraint on the emission region
size can be obtained from rapid variability, then all relevant
parameters can be uniquely determined \citep[e.g.,][]{tmg98}.
However, recent new discoveries by VERITAS with good
simultaneous or quasi-simultaneous multiwavelength coverage
provided strongly constraining SEDs that were difficult to
reproduce with such a model. 

\begin{figure*}
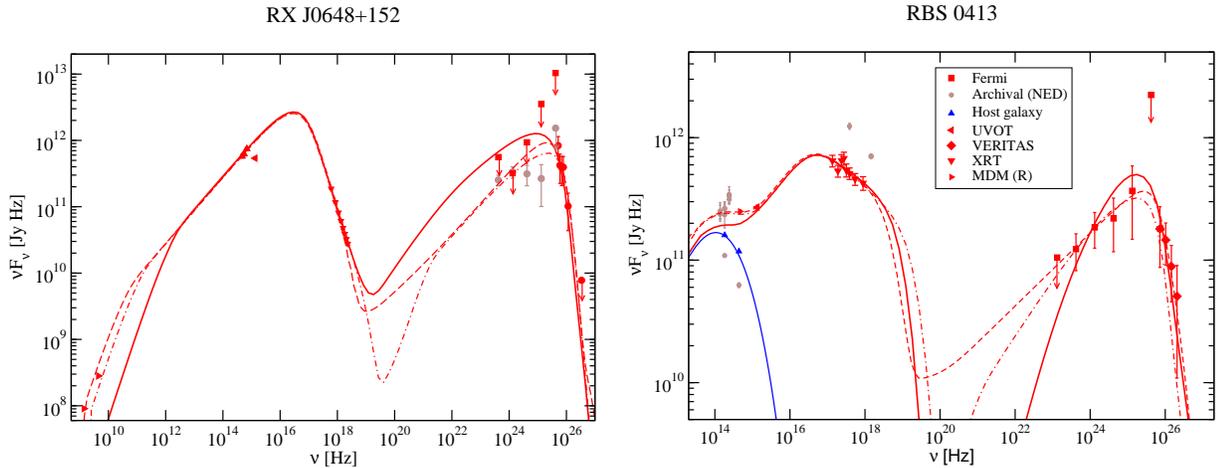

\centering
\includegraphics[width=0.46\textwidth]{fig2a.eps}
\hskip 0.3cm
\includegraphics[width=0.46\textwidth]{fig2b.eps}
\caption{\label{noSSCHBLs}Spectral energy distributions of the HBLs
RX J0648.7+1516 \citep[left;][]{aliu11} and RBS~0413 \citep[right;][]{aliu12}.
The {\it Fermi} upper limits for RX~J0648.7+1516 correspond to the 
$\sim 2$~months time window simultaneous with the VERITAS observations,
while the brown {\it Fermi} points show the 2-year average spectrum.
The solid curve indicates a pure leptonic SSC fit. The dashed dot-dashed
curve is a leptonic SSC + EC fit, while the dashed curve shows a fit
with the semi-analytical hadronic model described in Section \ref{hadronic}.
See text for details of the individual fits. The blue curve in the right
panel represents a phenomenological representation of the (thermal
dominated) host-galaxy emission. }
\end{figure*}

The HBL RX~J0648.7+1516 ($z = 0.179$) was detected as a VHE $\gamma$-ray
source by VERITAS in 2010 March -- April \citep{aliu11}. Simultaneous
multiwavelength coverage was provided by {\it Fermi}-LAT, {\it Swift} XRT
and UVOT and the MDM Observatory (optical B, V, R bands). The resulting
SED is shown in the left panel of Figure \ref{noSSCHBLs}. An attempt to
fit this SED with a pure SSC model clearly overshoots both the simultaneous
{\it Fermi} upper limits and the 2-year average spectrum. This is a
consequence of the apparent sharp spectral break between the {\it Fermi}
and VERITAS energy ranges, which is not consistent with the smooth
curvature of an SSC spectrum (resulting from the broad ranges of 
scattering electron energies and synchrotron seed photon frequencies, 
in combination with the gradual onset of Klein-Nishina effects). 
A much more satisfactory leptonic fit can be achieved when allowing 
for an EC contribution to the $\gamma$-ray emission. In the fit shown 
by the dot-dashed curve in the left panel of Fig. \ref{noSSCHBLs}, an 
external blackbody radiation field at a temperature of $T = 1000$~K
has been assumed, which could result from an infrared-emitting dust region 
around the AGN. Its luminosity corresponds to a flux of $\nu F_{\nu}^{\rm IR}
\lesssim 10^9$~Jy Hz, in agreement with its non-detection in the SED.

Alternatively, the hadronic model described above also provides a
satisfactory fit to the SED (the dashed curve in the left panel
of Fig. \ref{noSSCHBLs}).
The model requires protons to be accelerated up to $E_{\rm p, max} =
1.5 \times 10^{19}$~eV, with a total kinetic luminosity in relativistic
protons of $L_p = 2 \times 10^{45}$~erg/s. While hadronic models often 
suffer from extreme power requirements, this appears to be a very 
reasonable jet power. The magnetic field of $B = 10$~G is slightly
above equipartition with the energy content in relativistic protons.

The HBL RBS~0413 ($z = 0.19$) was detected by VERITAS in data accumulated 
between 2008 September and 2010 January. Contemporaneous multiwavelength 
coverage was provided by {\it Fermi}-LAT, {\it Swift} XRT and UVOT, 
and MDM (R-band). A leptonic SSC model fit (solid curve in the right
panel of Fig. \ref{noSSCHBLs}) produces a GeV $\gamma$-ray
spectrum significantly harder than indicated by {\it Fermi}-LAT,
and requires a magnetic field energy density a factor $\sim 6
\times 10^{-2}$ below equipartition. A much more satisfactory
representation of the entire SED (including the {\it Fermi}-LAT
spectrum), can be achieved by including an EC component with
a thermal blackbody radiation field at $T = 1.5 \times 10^3$~K.
This is shown by the dot-dashed curve in the right panel of
Figure \ref{noSSCHBLs}).
Parameters in almost exact equipartition between the magnetic field
and the relativistic electron population can be chosen, and the
postulated infrared radiation field corresponds to a flux of
$\nu F_{\nu}^{\rm IR} \sim 5 \times 10^8$~JyHz, consistent with 
the absence of direct evidence for it in the infrared spectrum
of the source. 

The dashed curve in the right panel of Fig. \ref{noSSCHBLs} shows a hadronic 
fit to the SED of RBS~0413, which also provides a better representation
than the pure SSC model. For this fit, protons need to be accelerated
to $E_{\rm p, max} = 1.6 \times 10^{19}$~eV with a kinetic power of
$L_p = 2 \times 10^{46}$~erg/s. The magnetic field of $B = 30$~G
is in almost exact equipartition with the energy content in
relativistic protons. 

Thus, both RX~J0648.7+1516 and RBS~0413 exhibit SEDs that disfavor
a pure leptonic single-zone SSC interpretation, while both an EC
model using a thermal infrared radiation field, and a hadronic
model provide convincing fits. Both EC and hadronic model fits 
can be achieved with with reasonable energy requirements and 
parameters close to equipartition between the magnetic field
and the dominant relativistic particle population.

\section{\label{inhomogeneous}Inhomogeneous jet models}

The complicated and often inconsistent variability features found
in blazars provide a strong motivation to investigate jet models
beyond a simple, spherical, one-zone geometry. The idea behind
phenomenological multi-zone models like the spine-sheath model
of \cite{tg08} or the decelerating-jet model of \cite{gk03} was
that differential relativistic motion between various emission
zones will lead to Doppler boosting of one zone's emission into
the rest frame of another zone. This can reduce the requirements
of extreme bulk Lorentz (and Doppler) factors inferred from simple
one-zone leptonic modeling of rapidly variable VHE $\gamma$-ray
blazars, such as Mrk~501 or PKS~2155-304, and has led to successful
model fits to the SEDs of those sources with much more reasonable
model parameters.

However, the models mentioned above do not treat the radiation
transport and electron dynamics in a time-dependent way and do
therefore not make any robust predictions concerning variability
and inter-band cross-correlations and time lags. In order to
address those issues, much work has recently been devoted to the
investigation of the radiative and timing signatures of shock-in-jet
models, which will be summarized in the following sub-section.

\subsection{\label{shock}Shock-in-jet models}

Early versions of shock-in-jet models were developed with focus
on explaining radio spectra of extragalactic jets, e.g., by
\cite{mg85}. Their application to high-energy spectra of blazars
was proposed by \cite{spada01}. Detailed treatments of the electron
energization and dynamics and the radiation transfer in a standing
shock (Mach disk) in a blazar jet were developed by
\cite{sokolov04,sm05,graff08}. The internal-shock model discussed 
in \cite{mimica04} and \cite{jb11} assumes that the central engine 
is intermittently ejecting shells of relativistic plasma at varying 
speeds, which subsequently collide. Such models have had remarkable 
success in explaining SEDs and time lag features of generic blazars.

\begin{figure}
\vskip 1cm
\centering
\includegraphics[width=0.48\textwidth]{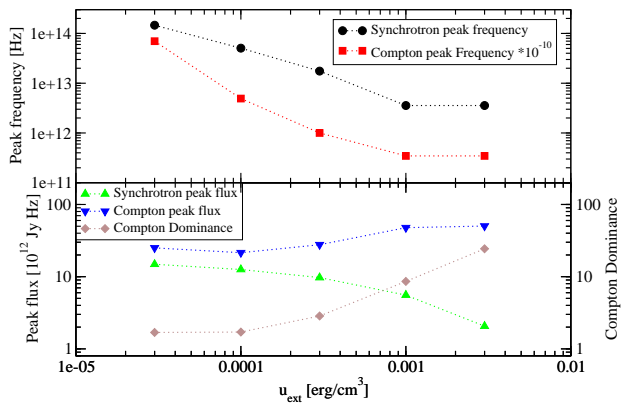}
\caption{\label{cs_sppar} Dependence of the SED characteristics of
time-averaged spectra from the intermal-shock model, on the external
radiation energy density $u_{\rm ext}$. }
\end{figure}

The realistic treatment of radiation transfer in an internal-shock
model for a blazar requires the time-dependent evaluation of retarded
radiation fields originating from all parts of the shocked regions
of the jet. The model system is therefore highly non-linear and can
generally only be solved using numerical simulations
(e.g., \cite{sokolov04,mimica04,graff08}). As the current
detailed internal-shock models employ either full expressions or
accurate approximations to the full emissivities of synchrotron
and Compton emission, a complete simulation of the time-dependent
spectra and light curves is time-consuming and does therefore
generally not allow to efficiently explore a large parameter
space. General patterns of the SED, light curves and expected
time lags between different wavelength bands have been demonstrated
for very specific, but observationally very poorly constrained,
sets of parameters. The most detailed treatment of the 
radiation-transfer aspect in the context of this models, including
all light-travel time effects, has been presented in \cite{chen11},
who employ a time-dependent multi-zone Monte-Carlo (for the radiation
transfer) and Fokker-Planck (for electron dynamics) code. 

In order to be able to explore timing signatures of the internal-shock
model for blazars on the basis of a broad parameter study, \cite{bd10} 
have developed a semi-analytical scheme for this model. The time- and 
space-dependent electron spectra, affected by shock acceleration behind 
the forward and reverse shocks, and subsequent radiative cooling, are 
calculated fully analytically. Taking into account all light travel 
time effects, the observed synchrotron and external-Compton spectra 
are also evaluated fully analytically, using a $\delta$-function 
approximation to the emissivities. The evaluation of the SSC emission 
still requires a two-dimensional numerical integration.

This semi-analytical model allowed the authors to efficiently scan a
substantial region of parameter space and discuss the dependence on
the characteristics of time-averaged SEDs, as well as cross-band
correlations and time lags. As an example, Fig. \ref{cs_sppar} shows
the dependence of the SED characteristics on the external radiation
field energy density, $u_{\rm ext}$. In the classical interpretation
of the blazar sequence, an increasing $u_{\rm ext}$ corresponds to
a transition from BL Lac spectral characteristics to FSRQ-like
characteristics. The decreasing synchrotron peak frequency and
increasing Compton dominance found in the parameter study are
reproducing this effect.

\begin{figure}
\vskip 1cm
\centering
\includegraphics[width=0.48\textwidth]{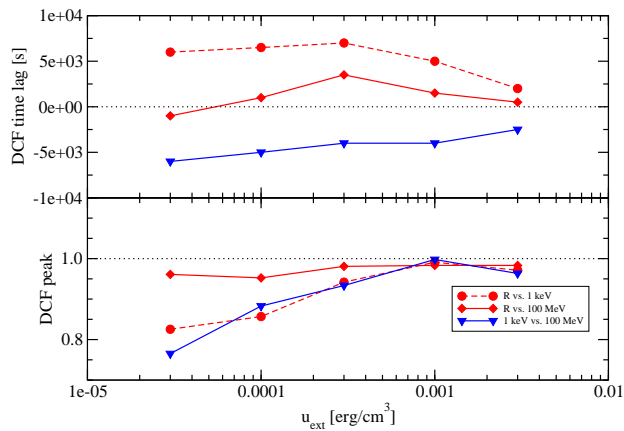}
\caption{\label{cs_dcfpar} Dependence on inter-band cross correlations
and time lags on the external radiation energy density. }
\end{figure}

From the internal-shock simulations, energy-dependent light curves
could be extracted. Using the standard Discrete Correlation Function
(DCF: \cite{ek88}) analysis, inter-band time lags could be extracted
for any set of parameters. Fig. \ref{cs_dcfpar} shows the dependence
of the inter-band time lags between optical, X-ray and \fermi\
$\gamma$-ray light curves (top panel) and the quality of the cross
correlation, characterized by the peak value of the DCF (bottom panel)
as a function of the external radiation energy density.

One of the most remarkable results of this study was that only slight
changes in physical parameters can lead to substantial changes of the
inter-band time lags and even a reversal of the sign of the lags. This
may explain the lack of consistency of lags even within the same source.

\bigskip
\begin{acknowledgments}
This work has been supported by NASA through Astrophysics Theory Program
grant NNX10AC79G and Fermi Guest Investigator grants NNX09AT82G and
NNX10AO49G. 
\end{acknowledgments}

\bigskip % extra skip inserted
%\begin{thebibliography}{9}   % Use for  1-9  references

\end{document}